# Spin-orbit interactions of light

K.Y. Bliokh[1], F.J. Rodríguez-Fortuño[2], F. Nori[1,3], and A.V. Zayats[2]

[1]*Center for Emergent Matter Science, RIKEN, Wako-shi, Saitama 351-0198, Japan*
[2]*Department of Physics, King's College London, Strand, London WC2R 2LS, UK*
[3]*Physics Department, University of Michigan, Ann Arbor, Michigan 48109-1040, USA*

Light carries spin and orbital angular momentum. These dynamical properties are determined by the polarization and spatial degrees of freedom of light. Modern nano-optics, photonics, and plasmonics, tend to explore subwavelength scales and additional degrees of freedom of structured, i.e., spatially-inhomogeneous, optical fields. In such fields, spin and orbital properties become strongly coupled with each other. We overview the fundamental origins and important applications of the main spin-orbit interaction phenomena in optics. These include: spin-Hall effects in inhomogeneous media and at optical interfaces, spin-dependent effects in nonparaxial (focused or scattered) fields, spin-controlled shaping of light using anisotropic structured interfaces (metasurfaces), as well as robust spin-directional coupling via evanescent near fields. We show that spin-orbit interactions are inherent in all basic optical processes, and they play a crucial role at subwavelength scales and structures in modern optics.

Light consists of electromagnetic waves that oscillate in time and propagate in space. Scalar waves are described by their intensity and phase distributions. These are *spatial (orbital)* degrees of freedom, common for all types of waves, either classical or quantum. In particular, propagation of a wave is associated with its phase gradient, i.e., the *wavevector* or *momentum*. Importantly, electromagnetic waves are described by *vector* fields[1]. Therefore, light also possesses intrinsic *polarization* degrees of freedom, which are associated with the directions of the electric and magnetic fields oscillating in time. In the quantum picture, the right-hand and left-hand circular polarizations, with the electric and magnetic fields rotating about the wavevector direction, correspond to two *spin* states of photons[2].

Recently there has been enormous interest in *spin-orbit interactions (SOI) of light*. These are striking optical phenomena where the spin (circular polarization) affects and controls the spatial degrees of freedom of light[3–6]. The intrinsic SOI of light originate from fundamental properties of the Maxwell equations[7,8] and are analogous to the spin-orbit interactions of relativistic quantum particles[2,9,10] and electrons in solids[11,12]. Therefore, fine SOI phenomena appear in all basic optical processes and require revisiting traditional approaches to many optical problems. To mention the most representative examples:

(i) A circularly polarized laser beam reflected or refracted at a dielectric interface does not propagate in the original plane but experiences a tiny transverse spin-dependent shift out of this plane. This is a manifestation of the *spin-Hall effect* of light[13–19].

(ii) Focusing of circularly-polarized light by a high-numerical-aperture lens results in the generation of a spin-dependent optical vortex (i.e., helical phase producing orbital angular momentum) in the focal field. This is an example of the *spin-to-orbital angular momentum conversion* in nonparaxial fields[20–27]. Breaking the cylindrical symmetry of a nonparaxial field also produces spin-Hall effect shifts[28–33].

(iii) A similar spin-to-vortex conversion occurs when a paraxial (collimated) beam propagates in an optical fiber[34] or along the optical axis of a uniaxial *anisotropic* crystal[35,36]. Moreover, properly designing anisotropic and inhomogeneous structures (e.g., metasurfaces or liquid crystals) allows considerable enhancement of the SOI effects and highly efficient spin-dependent shaping and control of light[37–45].

(iv) Any surface or waveguide mode possesses *evanescent* tails. Strikingly, coupling transversely-propagating spinning light to these evanescent tails results in a robust *spin-controlled unidirectional propagation* of the surface or waveguide modes[45–52]. This is a manifestation of the extraordinary transverse spin of evanescent waves[53,54] associated with the fundamental *quantum spin-Hall effect* of light[55].

Importantly, many of the above SOI phenomena appear at *subwavelength* scales. These effects become particularly conspicuous for *structured* waves with wavelength-scale inhomogeneities (including near-fields). That is why they remained largely neglected until recently. However, the rapid development of nano-optics, photonics, and plasmonics in the past two decades made the wavelength-scale processes crucially important for the manipulation and control of light. Tight focusing, interaction with nanoparticles and subwavelength structures, propagation in nanofibers and metamaterials, as well as near-field optics and plasmonics – all these research areas essentially involve structured fields, mix spatial and polarization degrees of freedom, and, hence, exhibit numerous SOI effects[56].

Below we overview the main SOI phenomena in optics, which were intensively studied in the past decade and are currently attracting rapidly growing interest. We examine the most important examples of the SOI of light in both paraxial and nonparaxial fields, natural media and artificial structures. In doing so, we always explain various manifestations of SOI using the same universal underlying concepts: *angular momenta* and *geometric phases*. Such unifying description provides a thorough understanding of SOI phenomena, explains all the main feature of their complex behavior in various systems, and illuminates their fundamental origin.

## Angular momenta and geometric phases

Two important fundamental concepts underpin the SOI of light: *optical angular momentum*[57–59] and *geometric (Berry) phases* for light[60–62]. Both these topics were intensively studied and reviewed in the past two decades, and here we only summarize basic aspects crucial for the understanding of SOI (see Boxes 1 and 2).

Light (or photons) carries momentum, which can be associated with its propagation direction and mean wave vector: $\mathbf{P} = \langle \mathbf{k} \rangle$ (hereafter we consider dynamical quantities per photon in $\hbar = 1$ units). Structured light also carries different kinds of angular momentum (AM). For paraxial (collimated) optical beams, AM can be decomposed in three separately-observable components: (i) *spin* AM (SAM), (ii) *intrinsic orbital* AM (IOAM), and (iii) *extrinsic orbital* AM (EOAM). These three types of optical AM can be associated with circular polarizations, optical vortices inside the beam, and beam trajectory, respectively. In addition to the momentum, these three AM are determined by the following key parameters: helicity $\sigma = \pm 1$ corresponding to the right-hand and left-hand circular polarizations, vortex quantum number $\ell$ that can take on any integer value, and transverse coordinates of the beam centroid $\mathbf{R} = \langle \mathbf{r} \rangle$ (see Box 1).

The interplay and mutual conversions between the three types of optical AM produce the optical SOI. Namely, the interaction between the SAM and EOAM results in a family of *spin-Hall effects*, i.e., helicity-dependent position or momentum of light. In turn, the coupling between the SAM and IOAM produces *spin-to-orbital AM conversions*, i.e., helicity-dependent optical vortices. Finally, the "orbit-orbit coupling" between IOAM and EOAM[63,64] causes the *orbital-Hall effects*: vortex-dependent shifts of optical beams. The latter phenomena are similar to the spin-Hall effects considered here and are mostly left outside of this review.

**Box 1. Angular momenta of light**

Paraxial optical beams can carry three types of angular momentum (AM)[57–59]. Firstly, the rotation of the electric and magnetic fields in a circularly-polarized beam produces spin AM (SAM). The SAM is aligned with the propagation direction (momentum $\mathbf{P} = \langle \mathbf{k} \rangle$) of the beam and is determined by the polarization helicity $\sigma \in (-1,1)$. The latter is the degree of circular polarization taking on $\pm 1$ values for right-hand and left-hand circular polarizations.

Secondly, optical vortex beams with helical phase fronts carry intrinsic orbital AM (IOAM). Akin to the SAM, this angular momentum is aligned with the momentum, and is determined by the vortex topological charge $\ell = 0, \pm 1, \pm 2, ...$ (i.e., the phase increment around the vortex core modulo $2\pi$).

Finally, beams propagating at some distance from the coordinate origin carry extrinsic orbital AM (EOAM). It is analogous to mechanical AM of a classical particle and is given by the cross-product of the transverse position of the beam center, $\mathbf{R} = \langle \mathbf{r} \rangle$, and its momentum $\mathbf{P}$. These parameters characterize the *trajectory* of the beam and may vary in inhomogeneus media.

The above optical angular momenta are schematically shown in Figure B1, and are described by the following expressions:

$$\mathbf{S} = \sigma \frac{\mathbf{P}}{P}, \qquad \mathbf{L}^{\text{int}} = \ell \frac{\mathbf{P}}{P}, \qquad \mathbf{L}^{\text{ext}} = \mathbf{R} \times \mathbf{P}.$$

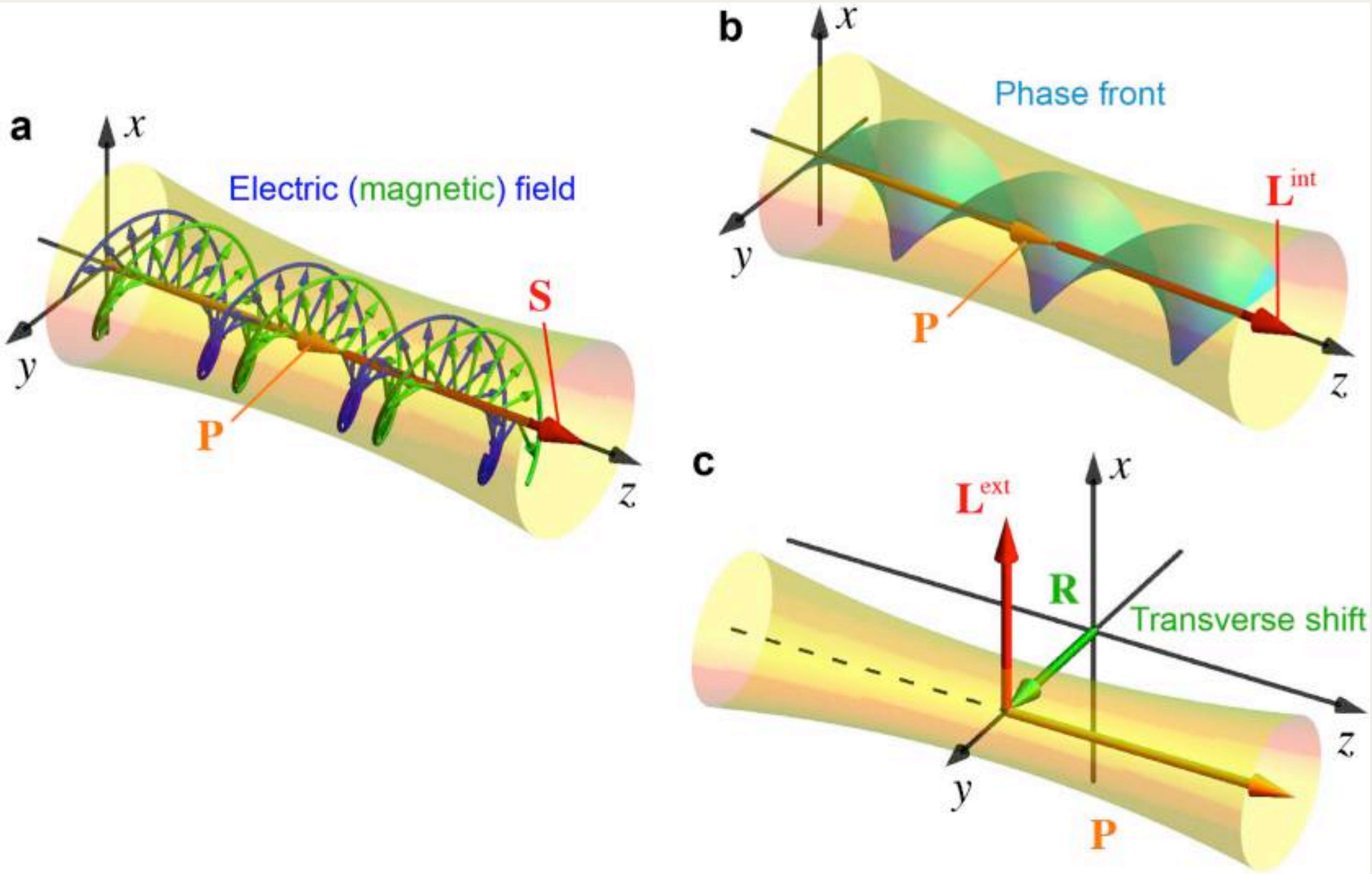

**Figure B1. Angular momenta of paraxial optical beams. (a)** Spin angular momentum for a right-hand circularly-polarized beam with $\sigma = 1$. **(b)** Intrinsic orbital angular momentum in a vortex beam with $\ell = 2$. **(c)** Extrinsic orbital angular momentum due to the propagation of the beam at some distance $\mathbf{R}$ from the coordinate origin.

Most importantly, the fundamental mechanisms underlying the spin-dependent deformation of optical fields are *geometric phases* (see Box 2). These can be explained as originating from the coupling between the SAM and rotations of the coordinate frames naturally determined in each particular problem. For example, rotating the transverse $(x,y)$ coordinates induce opposite phase shifts in the right-hand and left-hand circularly-polarized waves propagating along the $z$-axis. Such helicity-dependent phases underpin the spin-dependent shaping of light via two-dimensional anisotropic structures with varying orientation of the anisotropy axis[5,6,37–39,41,42]. A more sophisticated example of the geometric phase, which is inherent in free-space Maxwell equations, is related to three-dimensional variations of the direction of the wavevector $\mathbf{k}$ (and the SAM attached to it). Comparing phases of circularly-polarized waves propagating in different directions involves SO(3) rotations of coordinates and generates helicity-dependent geometric phases, which are described by the *Berry connection and curvature* on the sphere of directions in wavevector space[4,7,8,60–62,65] (Box 2). Such spin-redirection geometric phases underpin all intrinsic SOI phenomena, which take place in isotropic inhomogeneous media[3,4,7,13–19] and in nonparaxial free-space fields[4,8,20–33]. Note that the wavevector-dependent geometric phases occur for variations in *individual* wavevectors $\mathbf{k}$ in the Fourier spectrum of the field as well as for the evolution of the *mean* wavevector $\langle \mathbf{k} \rangle$ (momentum) of the whole beam.

**Box 2. Geometric phases**

Geometric phases in optics originate from the coupling between intrinsic angular momentum and *rotations of coordinates*. The simplest example for paraxial light is shown in Figure B1a. Circularly-polarized waves propagating in the $z$-direction and carrying SAM $\sigma\bar{\mathbf{z}}$ are characterized by the electric-field polarization vectors $\mathbf{E}^{\sigma} \propto \bar{\mathbf{x}} + i\sigma\bar{\mathbf{y}}$, where $\sigma = \pm 1$ and the overbars denote the unit vectors of the corresponding axes. Rotation of the coordinates by an angle $\varphi\bar{\mathbf{z}}$ (i.e., about the $z$-axis) induces helicity-dependent phases: $\mathbf{E}^{\sigma} \to \mathbf{E}^{\sigma} \exp(-i\sigma\varphi)$. This is the geometric phase $\Phi_{\rm G} = -\sigma\varphi$ given by the product of the SAM and the rotation angle.

This example allows a straightforward extension to the general case of nonparaxial (focused) light with arbitrary directions of propagation and rotations. If the wave carries SAM $\mathbf{S}$, and the coordinate frame experiences rotations with an angular velocity $\mathbf{\Omega}_{\zeta}$ (defined with respect to the parameter $\zeta$ which can be a coordinate or time), then the wave acquires a geometric phase during this $\zeta$-evolution: $\Phi_{\rm G} = -\int \mathbf{S} \cdot \mathbf{\Omega}_{\zeta}\, d\zeta$. This simple "dynamical" form[7,30] unifies the so-called Pancharatnam–Berry and spin-redirection types of geometric phase[60–62] and unveils its similarity with the rotational Doppler shift[131–133] and Coriolis effect[134–136].

An important example, underlying the SOI of light in isotropic media, is the geometric phase caused by variations of the wavevector direction, $\bar{\mathbf{k}} = \mathbf{k}/k$, in nonparaxial fields. The polarization of a plane wave in vacuum is always orthogonal to its wavevector: $\mathbf{k} \cdot \mathbf{E}(\mathbf{k}) = 0$. This is the *transversality condition*, which makes the polarization dependent on the wavevector and *tangent* to the $\bar{\mathbf{k}}$-sphere of directions in wavevector space (Fig. B2b). The geometric *parallel transport* of the polarization vector on the curved surface of this sphere reveals inevitable rotations between the transported vector and the global spherical coordinates, and, therefore, induces geometric phases in circularly-polarized waves. Using the helicity basis of circular polarizations $\mathbf{E}^{\sigma}(\mathbf{k})$ attached to the spherical coordinates $(\theta,\phi)$ in wavevector space, geometric-phase phenomena are described by the so-called *Berry connection* $\mathbf{A}^{\sigma}$ and *Berry curvature* $\mathbf{F}^{\sigma}$:[4,7,8,60–62,65]

$$\mathbf{A}^{\sigma}(\mathbf{k}) = -i\,\mathbf{E}^{\sigma *} \cdot (\nabla_{\mathbf{k}})\mathbf{E}^{\sigma} = -\frac{\sigma}{k}\cot\theta\,\bar{\boldsymbol{\phi}}, \qquad \mathbf{F}^{\sigma}(\mathbf{k}) = \nabla_{\mathbf{k}} \times \mathbf{A}^{\sigma} = \sigma\frac{\mathbf{k}}{k^3}.$$

Despite their geometrical origin, these unusual quantities act as an effective "vector-potential" and "magnetic field" in wave-momentum space, with helicity $\sigma$ playing the role of the "charge". Therefore, the Berry connection and Berry curvature immediately appear in all intrinsic SOI phenomena, such as the spin-Hall effect.

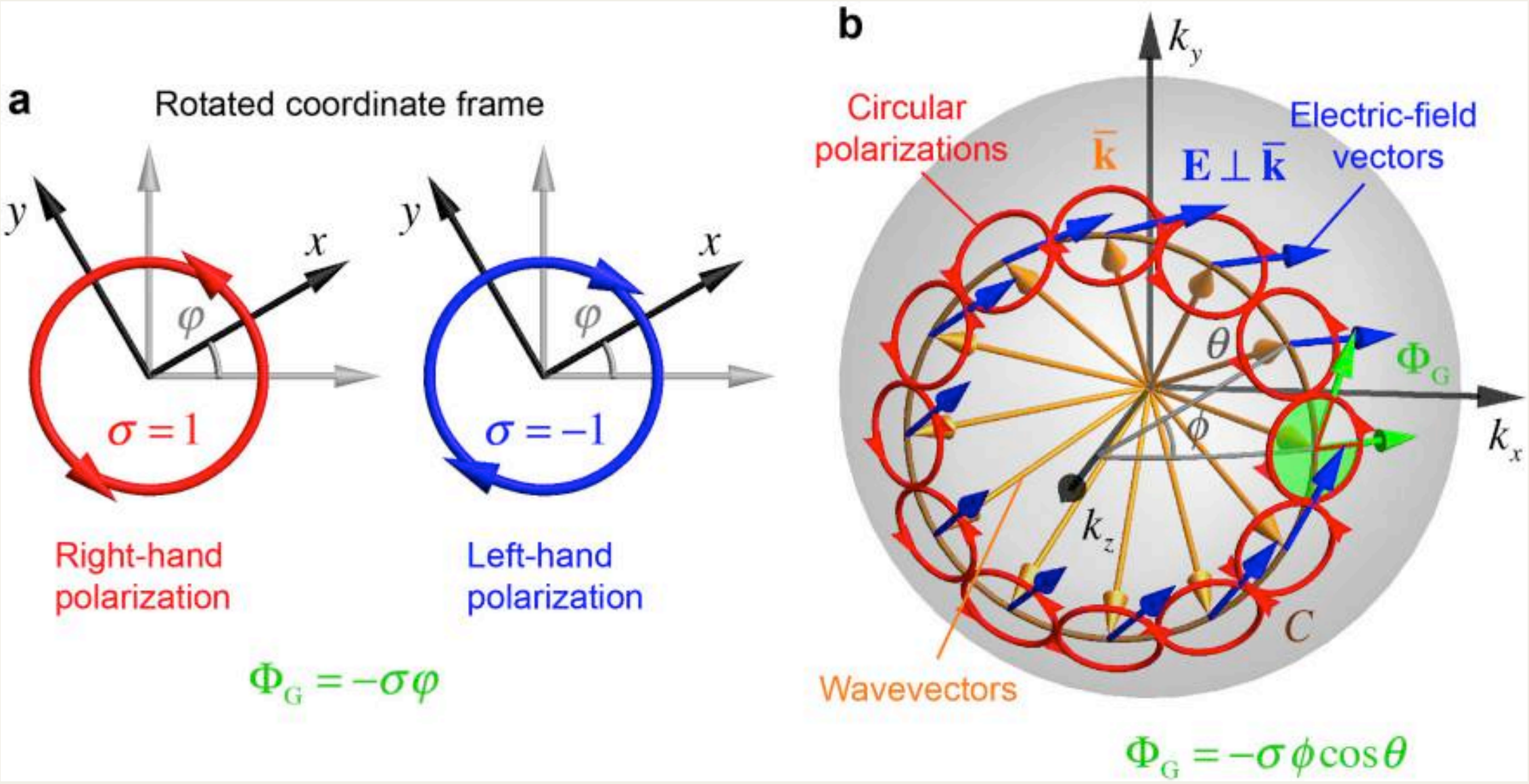

**Figure B2. Rotation-induced geometric phases. (a)** A two-dimensional rotation of the transverse coordinates induces a helicity-dependent phase shift in circularly-polarized light. **(b)** The three-dimensional variations in the wavevector direction involve non-trivial parallel transport of the polarization on the sphere of directions. The rotation of the transported vector with respect to spherical coordinates produces a helicity-dependent geometric phase difference between circularly-polarized waves propagating in different directions.

The Berry connection allows the comparison of phases of circularly-polarized waves propagating in different directions. Namely, variations of the wavevector along a contour $C$ on the $\bar{\mathbf{k}}$-sphere bring about the geometric phase $\Phi_G = \int_C \mathbf{A}^\sigma \cdot d\mathbf{k}$ (an analogue of the Aharonov–Bohm phase for the "vector-potential" $\mathbf{A}^\sigma$). In particular, traversing a $\theta = \text{const}$ contour, such as that shown in Fig. B2b, the right-hand and left-hand circularly polarized waves acquire opposite geometric phases $\Phi_G = -\sigma\phi\cos\theta$ (so the linear-polarization vector rotates by an angle $-\phi\cos\theta$). Remarkably, this exactly coincides with the "SAM-rotation coupling" expression $-S_z\phi$. For the whole loop, subtracting the $2\pi$ rotation of the $\bar{\boldsymbol{\phi}}$-coordinate, this yields the global phase $\Phi_{G0} = 2\pi\sigma(1-\cos\theta)$, determined by the solid angle enclosed by the contour[60–62].

## Spin-Hall effects in inhomogeneous media

The first important example of SOI occurs in the propagation of paraxial light in an *inhomogeneous isotropic* medium. It is well-known from geometrical optics that light changes its direction of propagation and momentum due to refraction or reflection at medium inhomogeneities. However, the trajectory of an optical beam is independent of the polarization in traditional geometrical optics in the absence of anisotropy[66]. This is because geometrical optics neglects all wavelength-scale phenomena, which become important for modern nano-optics.

Going beyond the geometrical-optics approximation and considering the wavelength-order corrections to the evolution of light introduces polarization-dependent perturbations of the light trajectory coming from the intrinsic SOI in Maxwell equations[7].

Let us consider the propagation of light in a gradient-index medium with refractive index $n(\mathbf{r})$. The smooth trajectory of a light beam in such medium can be described by the mean coordinates $\mathbf{R}$ and the momentum $\mathbf{P}$, which vary with the trajectory length $\tau$. Considering "semiclassical" (i.e., wavelength-order) corrections to this "mechanical" formalism, the trajectory of light in a gradient-index medium is described by the following equations of motion[3,7,13,14,17,67]:

$$\dot{\mathbf{P}} = \nabla n(\mathbf{R}), \qquad \dot{\mathbf{R}} = \frac{\mathbf{P}}{P} - \frac{\sigma}{k_0}\frac{\mathbf{P}\times\dot{\mathbf{P}}}{P^3}. \tag{1}$$

Here the overdot stands for the derivative with respect to $\tau$, $k_0 = \omega / c$ is the vacuum wavenumber, and we used dimensionless momentum $\mathbf{P} = \langle \mathbf{k} \rangle / k_0$. The last term in Eq. (1) describes the transverse spin-dependent displacement of the trajectory, i.e., the *spin-Hall effect of light* (Fig. 1a). This effect was originally called "optical Magnus effect"[3]. Later, it was shown that the helicity-dependent term in Eq. (1) can be considered as a "Lorentz force" produced by the Berry curvature $\mathbf{F}^{\sigma}(\mathbf{P})$ acting in momentum space[7,13,14,17] (Box 2). The Berry connection and curvature act as a geometry-induced "vector-potential" and "magnetic field" in momentum space, revealing the *geometrodynamical* nature of the SOI of light. In doing so, the Berry connection underlies the evolution of the polarization along the curvilinear trajectory, which obeys the parallel-transport law and is described by the geometric phases $\Phi_G = \int \mathbf{A}^{\sigma}(\mathbf{P}) \cdot d\mathbf{P}$ for the two helicity components (Fig. 1a)[7,60–62]. Measurements of this polarization evolution in coiled optical fibers was one of the first observations of the Berry phase in physics[68,69].

The spin-Hall effect and the equations of motion (1) for spinning light are completely analogous to those for electrons in condensed-matter[70] and high-energy[10] systems. While the electron's momentum is driven by an applied electric field, in optics the refractive-index gradient plays the role of an external driving force. Strikingly, the spin-Hall effect shows that an *isotropic* inhomogeneous medium exhibits *circular birefringence*. However, in contrast to anisotropic media, this birefringence is determined solely by intrinsic properties of light, namely, by its SAM. Moreover, the helicity-dependent shift of the trajectory is intimately related to the fundamental conservation of the total angular momentum of light. Indeed, for spherically-symmetric profiles $n(\mathbf{r})$, equations (1) possess the corresponding integral of motion[14] $\mathbf{J} = \mathbf{R}\times\mathbf{P} + \sigma\frac{\mathbf{P}}{P} = \mathbf{L}^{\text{ext}} + \mathbf{S} = \text{const}$.

Figures 1a,b show an example of the spin-Hall effect measured for the helical trajectory of light inside a glass cylinder[17]. The difference between the positions of the right-hand and left-hand circularly-polarized beams ($\sigma = \pm 1$) achieved several wavelengths, which was accumulated along several coils of the trajectory. This experimental observation has a fundamental importance for physics of relativistic spinning particles[10,67]. Indeed, direct measurements of analogous spin-dependent electron trajectories are far beyond the current experimental capabilities, and only indirect measurements of the spin-Hall effect are possible in condensed-matter physics[71].

Equations (1) describe a "macroscopic" picture of the spin-Hall effect, which contains only the *mean* beam parameters. What causes this unusual effect on the "microscopic" level of *individual* plane waves forming the beam? This can be understood considering another example of the spin-Hall effect.

Instead of a gradient-index medium, we now consider refraction or reflection of a paraxial beam at a sharp interface between two isotropic media. This problem is described by Snell's law and the Fresnel equations[1]. However, these equations are valid for a *single plane wave* impinging at the interface. At the same time, a finite size beam consists of multiple plane waves with slightly different wavevectors **k** (Fig. 1c), so that they have slightly different planes of incidence entering the Fresnel reflection/refraction equations. Let the $z$-axis be directed along the normal to the interface, and the incident-beam momentum lie in the $(x,z)$-plane with a polar angle of incidence $\theta$, i.e., $\langle k_y \rangle = 0$. Then, the planes of incidence for individual plane waves in the beam are rotated by the azimuthal angle $\phi = \bar{k}_y / \sin\theta$ about the $z$-axis (Fig. 1c) and, hence, induce *geometric phases* $\Phi_{\rm G}(k_y) = S_z \phi = \sigma \bar{k}_y \cot\theta$ for the circularly-polarized waves (Box 2)[18,72]. The $k_y$-gradient of this geometric phase determines a typical beam shift along the $y$-axis, i.e., out of the plane of incidence. Taking into account Fresnel coefficients of the interface and similar geometric phases for the reflected/refracted beams, one can obtain accurate equations for the spin-dependent shifts of these beams[15,18,72,73]. In the simplest case of total reflection from the interface, the reflected beam acquires the helicity-dependent shift

$$Y' = -\frac{\sigma + \sigma'}{k} \cot\theta \,, \tag{2}$$

where $\sigma'$ and $Y'$ are the helicity and centroid position of the reflected beam, and $Y = 0$ for the incident beam.

The transverse shift (2) is known as the Imbert–Fedorov shift, which has been predicted and observed a long time ago for total internal reflection of light[74,75]. However, studies of the Imbert–Fedorov effect were full of controversies. Only recently, correct theoretical calculations[15,72] and definitive measurements[16] have elucidated its nature as a SOI effect. (Note the close similarity between expression (2) and the Berry connection in Box 2[14].) Figure 1d shows measurements[16] of the spin-Hall splitting between the right-hand and left-hand circularly-polarized components in a linearly polarized beam refracted at the air-glass interface. Using the "quantum weak measurements" technique with near-orthogonal input and output polarizers[16,76,77], an amazing Angstrom accuracy was achieved.

Akin to equations (1), the spin-Hall shift (2) is intimately related to the interplay between the SAM and EOAM of the beams induced at the interface[15,72,78,79]. Namely, this shift ensures the conservation of the $z$-component of the total AM between the incident and reflected beams: $S_z = S'_z + L_z^{\rm ext\prime}$, where $S_z = \sigma\cos\theta$, $S'_z = -\sigma'\cos\theta$, and $L_z^{\rm ext\prime} = -Y' k \sin\theta$. Interestingly, the SOI of light at sharp interfaces also causes transverse polarization-dependent *deflections* (i.e., *momentum* shifts) of the reflected or refracted beams away from the angle predicted by geometrical optics[18,72,73,80]. Figure 1e shows the spin-Hall effect and images of the polarization-dependent coordinate and momentum shifts generated at the "refraction" of the incident beam of light into the surface plasmon-polariton beams propagating along a metal film[19]. Typically-subwavelength shifts are amplified to the beam-width scale using the "quantum weak measurements"[16,76,77].

Spin-Hall effects are ubiquitous to any optical interfaces. They have been measured for interfaces with metals[81], uniaxial crystals[82], and semiconductors[83], as well as at nanometal films[84], graphene layers[85], and metasurfaces[86]. The spin-Hall shifts exhibit an interesting anomaly near the Brewster angle[15,77,87,88] and a fine interplay with the Goos-Hanschen (in-plane) shifts[18,73,76,77,89]. Since every optical device and component operates with finite-size beams and not plane waves, the spin-Hall effects are always present at optical interfaces and inevitably affect the field distribution on the wavelength scale. On the one hand, they have to be taken into account as inevitable SOI-induced aberrations. On the other hand, "quantum weak measurement" amplification and dependence of the spin-Hall shifts on the material parameters allow employing the spin-Hall effect for precision metrology[84,85].

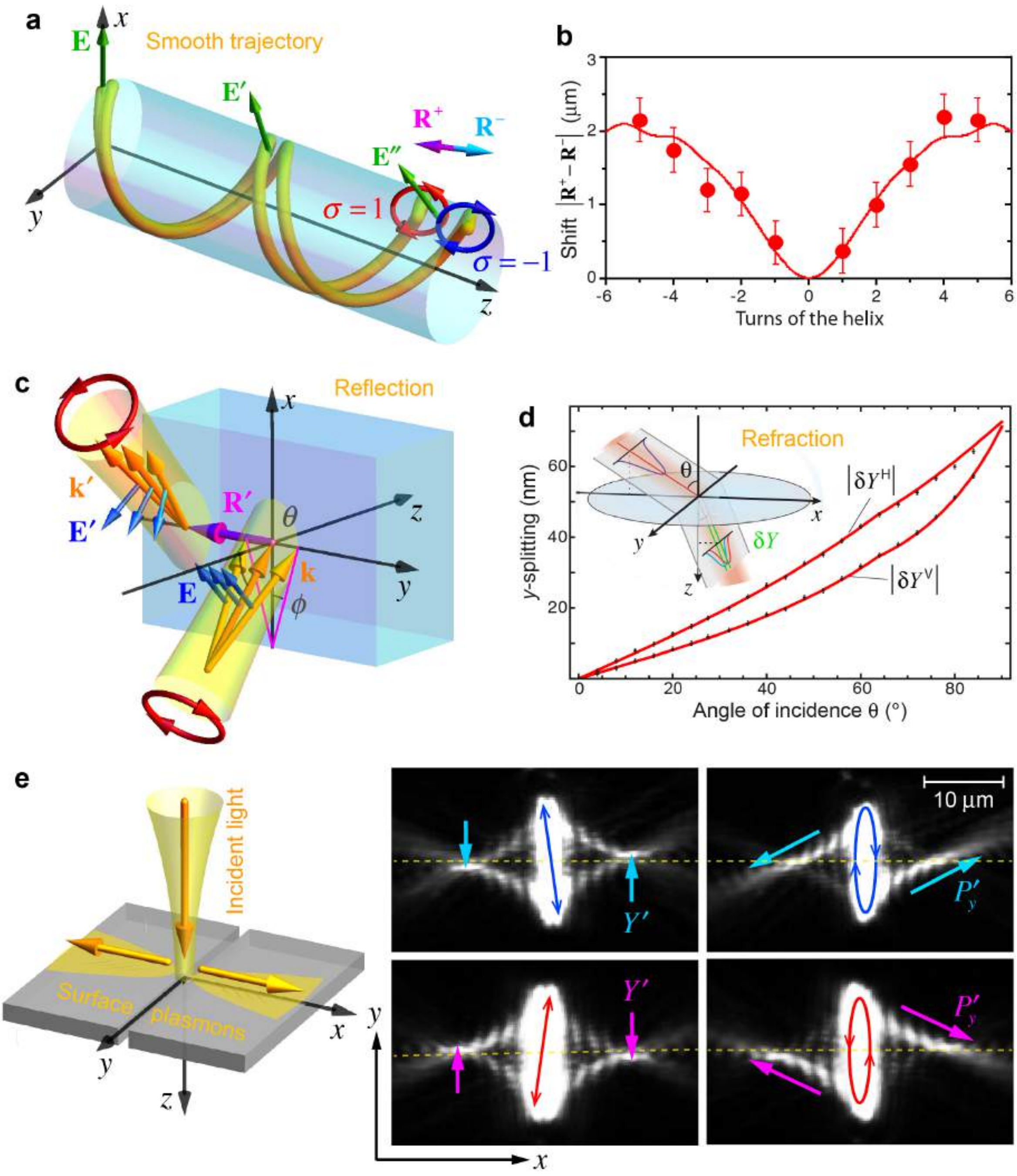

**Figure 1. Spin Hall effects for paraxial beams in inhomogeneous media. (a)** Propagation of light along a curvilinear trajectory causes transverse spin-dependent deflection produced by a "Lorentz force" from the Berry curvature (1). **(b)** Measurements of this spin-Hall effect for a helical light trajectory[17]. **(c)** A similar spin-dependent transverse shift (2) occurs in the beam reflection or refraction at a planar interface. The spin Hall effect is produced by $k_y$-dependent geometric phases acquired by different plane waves in the beam spectrum, which propagate in different planes (marked by azimuthal angles $\phi$). The spin-Hall shift generates extrinsic orbital AM (Fig. B1c) and provides for the AM conservation in the system. **(d)** Precision "quantum weak measurements" of the spin-Hall splitting in a linearly-polarized beam refracted at the air-glass interface[16]. **(e)** Observation of the spin-Hall shifts in both coordinate and momentum using the weak-measurement approach and "refraction" of the $z$-propagating light into $x$-propagating surface plasmon-polariton beams[19].

Thus, optical spin-Hall effects originate from the interaction between the SAM and EOAM, leading to mutual interrelations between the polarization and trajectory of light. A quite similar interaction between the IOAM and EOAM (vortex and trajectory) occurs for vortex beams in inhomogeneous media[18,64,90–95]. In this case, beams experience $\ell$-dependent transverse shifts at the medium inhomogeneities. These can be regarded as the "orbital-Hall effect" and "orbit-orbit interactions" of light.

## Spin-orbit interactions in nonparaxial fields

The above examples of the spin-Hall effect in isotropic media are based on intrinsic SOI properties which require variations of wavevectors in the field spectrum. This hints that the SOI may naturally be enhanced in nonparaxial fields: e.g., tightly focused by high-numerical-aperture lenses or scattered by small particles (Figs. 2a,d). Under such circumstances, the fields become inhomogeneous at the wavelength scale, and the SOI effects can strongly affect the field distributions.

Remarkably, the SOI manifest itself even in *free-space* nonparaxial fields. Consider, for example, focused circularly-polarized vortex beams carrying spin and orbital intrinsic AM. The simple association of the SAM and IOAM with the polarization and vortex (applicable for paraxial beams in Box 1), respectively, is not valid anymore. For nonparaxial beams, consisting of circularly-polarized plane waves with the wavevectors forming a cone with an opening angle $\theta$ (like in Fig. B2b), the SAM and IOAM become[8,27,96–98]

$$\mathbf{S} = \sigma \cos\theta \frac{\mathbf{P}}{P}, \quad \mathbf{L} = \left[\ell + \sigma\left(1-\cos\theta\right)\right]\frac{\mathbf{P}}{P}. \tag{3}$$

The total intrinsic AM of the beam is preserved: $J_z = S_z + L_z = \sigma + \ell$, so that Eqs. (3) can be interpreted as if a part of the SAM was transferred to the IOAM. This is another fundamental manifestation of the SOI: the *spin-to-orbital AM conversion*. Part of the *orbital* AM becomes *helicity*-dependent, i.e., a helicity-dependent vortex should appear even in beams with $\ell = 0$ (Fig. 2b)[22–24,27]. Importantly, this effect is closely related to the *geometric phase* between the azimuthally-distributed wavevectors $\mathbf{k}$ in the beam spectrum (Fig. B2b)[8,27]. Using the global geometric phase $\Phi_{\rm G0}$ between these wavevectors (Box 2), the converted part of the AM can be written as $\Delta L = \sigma \Phi_{\rm G0}/2\pi$. For the largest aperture angle $\theta = \pi/2$, $\Phi_{\rm G0} = 2\pi$, and the conversion efficiency reaches 100%, i.e., all the paraxial SAM becomes part of IOAM[25].

To understand the origin of the spin-to-orbital AM conversion, note that focusing by a high-NA lens rotates the wavevector (rays) of the incoming collimated beam in the meridional planes and generates a conical $\mathbf{k}$-distribution in the focused field, Fig. 2a. This is accompanied by rotations of the local polarization vectors $\mathbf{E}$ attached and orthogonal to each $\mathbf{k}$. Notably, this polarization evolution (described by the Debye–Wolf approach[100]) represents *parallel transport* on the $\bar{\mathbf{k}}$-sphere of directions (Fig. B2b) from the North pole $\theta = 0$ (incoming light $\mathbf{E}$) to $\theta \neq 0$ (focused field $\mathbf{E}'$). In the global basis of the circular ($\bar{\mathbf{x}} + i\sigma\bar{\mathbf{y}}$)-polarizations and longitudinal $\bar{\mathbf{z}}$ -component, this 3D rotational transformation of the electric field is described by the following unitary matrix[27]:

$$\mathbf{E}' = \begin{pmatrix} a & be^{-2i\phi} & \sqrt{2ab}\,e^{-i\phi} \\ -be^{2i\phi} & a & \sqrt{2ab}\,e^{i\phi} \\ -\sqrt{2ab}\,e^{i\phi} & -\sqrt{2ab}\,e^{-i\phi} & a-b \end{pmatrix} \mathbf{E}, \tag{4}$$

where $a = \cos^2(\theta/2)$ and $b = \sin^2(\theta/2)$. In equation (4), the off-diagonal elements contain the azimuthal *vortex* factors and are responsible for the AM conversion. Due to these elements, the incoming circularly-polarized light with helicity $\sigma$ acquires an oppositely-polarized component with helicity $-\sigma$ and vortex factor $b\exp(2i\sigma\phi)$, and also a longitudinal $z$-component with the vortex $\sqrt{2ab}\exp(i\sigma\phi)$. For small apertures, the longitudinal component plays the leading role in the AM conversion. These helicity-dependent vortex components produce the helicity-dependent IOAM (3) in the focused field (Fig. 2a)[8,24,27,99].

The presence of the helicity-dependent vortices and IOAM in focused light was observed using probe particles interacting with the focal field[22,23,99], Fig. 2b. The particles experienced transverse *orbital* rotation around the beam axis, which is characteristic of optical vortices[101,102], with the sense of a rotation determined by the *helicity* of the incoming wave, which had no vorticity prior to focusing. Such mechanical manifestations of the SOI can play an important role in optofluidics and optical manipulations using nonparaxial light.[103]

Notably, the above AM conversion immediately reveals itself in the helicity-dependent *intensity* distributions of the focused fields. Namely, the mean radius of a focused vortex beam is determined by its $\sigma$-dependent IOAM value: $R \sim |L_z| / k\sin\theta$ [8,27]. Due to this effect, a beam with parallel SAM and IOAM becomes more strongly focused than a similar beam with anti-parallel SAM and IOAM. The most striking manifestation of this effect appears for vortex beams with $|\ell| = 1$, Fig. 2c. According to the transformation (4), the $z$-component of the field has the vortex charge $\ell + \sigma$. Therefore, for $\ell\sigma = 1$ this component represents a charge-2 vortex with *vanishing* intensity in the beam center, while for $\ell\sigma = -1$ this is a charge-0 vortex with *maximum* intensity in the center. This helicity-dependent switching of the central intensity was observed in experiments[20,25] and employed for spin-controlled transmission of light via chiral nano-apertures[104] and AM-induced circular dichroism in non-chiral structures[105].

The SAM–IOAM coupling in nonparaxial optical fields is largely independent of how the field was generated. Instead of high-NA focusing, one can consider dipole Rayleigh scattering by a small particle, which generates a similar conical distribution of the outgoing wavevectors, Fig. 2d. The electric-field transformation in dipole scattering to the far-field direction $\bar{\mathbf{k}} = \bar{\mathbf{r}}$ is given by[1] $\mathbf{E}' \propto -\bar{\mathbf{r}} \times (\bar{\mathbf{r}} \times \mathbf{E})$, and it can be written in a matrix form very similar to Eq. (4).[27] Therefore, the spin-to-orbital AM conversion appears in the scattering of circularly-polarized light[21,26,27,106–108]. Since both focusing and scattering produce strong SOI, these phenomena play an important role in high-resolution optical microscopy and imaging of scattering processes[31,109]. This can be seen in the Stokes polarimetry of the paraxial field at the output of the imaging system. Superposition of the original $\sigma$-polarized state and the converted ($-\sigma$)-polarized state with the $\exp(2i\sigma\phi)$ vortex generates characteristic "four-petal" patterns in the first and second Stokes parameters $\mathcal{S}_1$ and $\mathcal{S}_2$, Fig. 2e. This effect was observed in systems of different nature and scales: diffusive backscattering from microparticle suspensions[109–111], scattering by liquid-crystal droplets[40], and dipole nanoparticle scattering[31].

The spin-to-orbital AM conversion can be interpreted as an azimuthal spin-Hall effect in cylindrically-symmetric fields[24,26,112]. Breaking this symmetry results in a pronounced spin-Hall effect in the direction orthogonal to the symmetry-breaking axis[8,28–33]. For example, illuminating only the $x > 0$ half of a high-NA lens results in the subwavelength transverse shift of the focal spot[8,28–30]: $Y' \propto \sigma / k$. Moreover, a high-NA microscope with a dipole-scatterer specimen allows a dramatic inversion of the spin-Hall effect scale[31]. Instead of subwavelength shifts caused by helicity switching (Fig. 1), subwavelength $x$-displacements of the particle cause giant macroscopic $y$-redistribution of the SAM density (i.e., the third Stokes parameter $\mathcal{S}_3$) in the exit pupil, Fig. 2f. Analogous "orbital-Hall effect", i.e., the $\ell$-dependent transverse redistribution of the intensity, can be seen in the asymmetric scattering of vortex beams[113].

These examples show that SOI crucially affects the distributions and properties of every instance of nonparaxial light, including fields interacting with subwavelength structures. As such, SOI phenomena inevitably emerge in numerous nano-optical, plasmonic, and metamaterial systems, all crucially involving subwavelength scales and structures[56].

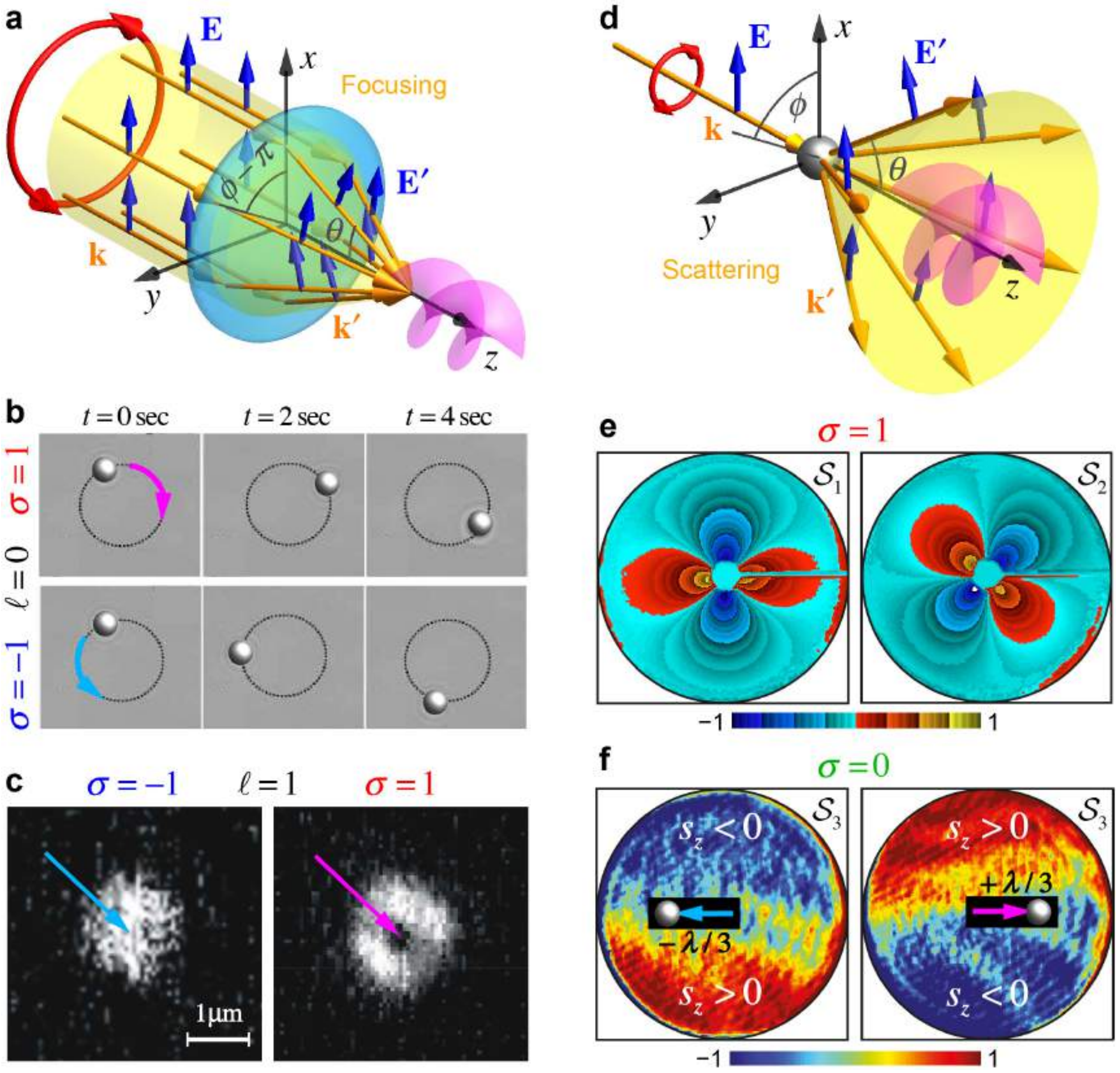

**Figure 2. Spin-orbit interactions in nonparaxial light. (a)** Tight focusing of a paraxial wave generates a conical wave-vector distribution (cf., Fig. B2b). The resulting 3D field has components with helicity-dependent vortices (i.e., intrinsic orbital AM). This is the spin-to-orbital AM conversion, Eqs. (3) and (4). **(b)** Experimental observation[99] of the helicity-dependent vortex and orbital AM in a focused field via the helicity-dependent orbital motion of a probe particle. **(c)** Manifestation of the spin-to-orbital AM conversion in the helicity-dependent intensity of the focused field[20]. Tightly focused beams with $\ell\sigma = 1$ and $\ell\sigma = -1$ have zero and maximal intensity in the center, respectively. **(d)** Rayleigh scattering by a small dipole particle produces a spherical redistribution of the field and the AM conversion, similar to the focusing case. **(e)** The spin-to-orbital AM conversion is clearly seen in the imaging and polarimetry of scattering processes as "four-petal" patterns in the Stokes parameters $\mathcal{S}_{1,2}$. Here experimental figures[109] for the diffusion-backscattered of light from a particle suspension are shown. **(f)** Giant spin-Hall effect induced by the breaking of cylindrical symmetry in the system. Subwavelength $x$-displacements of a Rayleigh nanoparticle in a high-NA imaging system produce a macroscopic $y$-separation of the spin AM density (the third Stokes parameter $\mathcal{S}_3$) in the linearly-polarized light[31].

## Spin-orbit interactions produced by anisotropic structures

Until now we considered only “intrinsic” SOI effects, which originate from fundamental properties of the Maxwell equations and not related to specific media. Such SOI phenomena are quite robust with respect to perturbations and particular details of the media. Another class of SOI effects can be induced by particular properties and symmetries of the medium. These “extrinsic” effects emerge in anisotropic media and artificial structures, including metamaterials, and thus can be designed to achieve one or another functionality. Strong anisotropy and inhomogeneities allow efficient control of the polarization degrees of freedom as well as controllable shaping of the intensity and phase distributions. In such media, strong SOI can be achieved even with $z$-propagating paraxial light interacting with $(x,y)$-planar structures. In this case, varying the orientation of anisotropic scatterers produces simple 2D geometric phases (Fig. B2a) leading to SOI.

Let us consider light transmission through a planar anisotropic element, Fig. 3a. For simplicity, we assume a transparent retarder providing a phase $\delta$ with the anisotropy axis oriented at an angle $\alpha$ in the $(x,y)$-plane. In the coordinates attached to the anisotropy axis, the evolution of light is described by the transmission Jones matrix $T = \mathrm{diag}\left(e^{i\delta/2}, e^{-i\delta/2}\right)$. Performing a rotation by the angle $\alpha$ to the laboratory coordinate frame and also writing this matrix in the helicity basis of right-hand and left-hand circular polarizations, the Jones-matrix transformation of the wave polarization becomes[5,6,37–39,114]

$$\mathbf{E}' = \begin{pmatrix} \cos\dfrac{\delta}{2} & i\sin\dfrac{\delta}{2}e^{-2i\alpha} \\ i\sin\dfrac{\delta}{2}e^{2i\alpha} & \cos\dfrac{\delta}{2} \end{pmatrix}\mathbf{E}. \qquad (5)$$

Here the off-diagonal elements with phase factors $\exp(\mp 2i\alpha)$ originate from geometric phases induced by the rotation of coordinates (Fig. B2a). For the half-wave retardation $\delta = \pi$, the matrix (5) becomes off-diagonal and describes the transformation of the $\sigma = \pm 1$ circularly-polarized light into the opposite polarization $\sigma' = \mp 1$, with the geometric phase difference $\Phi_G = -2\sigma\alpha$.

The off-diagonal geometric-phase elements of Eq. (5) offer helicity-dependent manipulation of light using the orientation of the anisotropy axis. In particular, anisotropic subwavelength gratings with *space-variant* orientation $\alpha = \alpha(x,y)$ have been employed for the geometric-phase-induced shaping of light[5,6,37–39,41,114–117]. Two most important cases of such “metasurfaces” are shown in Figs. 3b,c.

Let the orientation of the anisotropy axis change linearly with one of the coordinates: $\alpha = \alpha_0 + qx$, Fig. 3b. In this case, for a half-wave retardation, the $\sigma$-polarized light is converted into light of opposite helicity and also acquires the helicity-dependent geometric-phase gradient $\Phi_G = -2\sigma qx$. This phase gradient produces a transverse helicity-dependent component in the momentum (wavevector) of light: $P_x' = -2\sigma q$. Thus, the $x$-variant anisotropic structure deflects right-hand and left-hand polarized beams in opposite $x$-directions[5,37,42,114]. This can be considered as the anisotropy-induced *spin-Hall effect of light*. While in the intrinsic spin-Hall effect (Fig. 1) the coordinate shift is caused by the wavevector gradient of the geometric phase Fig. B2b, here the momentum shift is generated by the coordinate gradient of the geometric phase Fig. B2a. This extrinsic spin-Hall effect generated by the space-variant anisotropic elements allows complete spatial separation of the two spin states of light: a linearly-polarized light with $\sigma = 0$ is transformed into two well-separated $\sigma' = \pm 1$ beams propagating in different directions (Fig. 3b)[37]. Moreover, if the element converts the transmitted beam into the $x$- propagated surface-plasmon waves, then the two spin components propagate in *opposite* directions[42]. This provides a helicity-controlled directional coupler, which can be implemented in a variety of photonics platforms.

Assume now that the anisotropy-axis orientation varies linearly with the *azimuthal* coordinate $\varphi$ in the $(x, y)$-plane: $\alpha = \alpha_0 + q\varphi$ (Fig. 3c). Here $q = 0, \pm 1/2, \pm 1, ...$ is a half-integer number, and the structure has a direction singularity at the coordinate origin. In this case, the anisotropic half-wave plate reverses the helicity and generates the azimuthal geometric-phase gradient: $\Phi_G = -2\sigma q \varphi$. This means that the transmitted beam becomes a *vortex* beam with topological charge $\ell' = -2\sigma q$.[5,6,38,39,117] In other words, a *spin-to-orbital AM conversion* takes place. Such azimuthal anisotropic structures (also called *q*-plates) offer efficient spin-controlled converters and generators of optical vortex beams carrying IOAM. Note that the $q = 1$ anisotropic plate is rotationally symmetric with respect to the $z$-axis (Fig. 3c). In this case, the $z$-component of the total AM is conserved: $\sigma = \sigma' + \ell'$, and the Jones matrix (5) resembles the transverse $(x, y)$-sector of the matrix (4). Very similar conversions of the SAM into IOAM with $\ell' = -2\sigma$ occur in all cylindrically-symmetric systems with effective anisotropy between the radial and azimuthal polarizations. Examples include the propagation of light along the optical axis of a uniaxial crystal[35,36,118], in cylindrical optical fibers[34], as well as the focusing and scattering in rotationally-symmetric systems with paraxial input and output[31,40]. In the generic $q \neq 1$ case, the rotational symmetry is absent, there is no AM conservation for light, and part of the optical AM is transferred to the medium[119,120].

The above examples demonstrate that inhomogeneous anisotropic planar structures provide a highly efficient tool for spin-dependent shaping and control of light. Recently there has been an enormous interest to such structures, which can be considered as planar metamaterials, i.e., *metasurfaces*[121]. In the above examples, we implied two-scale structures with subwavelength gratings providing local anisotropy and inhomogeneity larger than the wavelength (but smaller than the beam size). If typical scales of the structure are comparable with the wavelength, such inhomogeneities can considerably modify the eigenmodes and spectral properties of light. Such structures can couple light to surface plasmon-polaritons and control properties of these surface waves. In particular, chiral structures can generate a spin-dependent intensity plasmonic distribution with vortices[25,122,123], while periodic crystal-like structures with breaking spatial-inversion symmetry result in spin-dependent spectra of photonic quasi-particles[43–45,124]. The latter case is entirely analogous to the spin-dependent splitting of electron energy levels in solids with SOI[11,12]. Figure 3d shows an example[44] of such plasmonics metasurface with the experimentally-measured spin-polarized dispersion. The different spin states of the incident light are coupled to different propagation directions of surface plasmons, depending on the frequency and orientation of the plasmonic crystal. Thus, the SOI of light at metasurfaces paves the avenue to spin-controlled photonics: an optical analogue of solid-state spintronics.

## Spin-direction locking via transverse spin in evanescent waves

After discussing artificial structures, we are coming back to fundamental intrinsic properties of light. Recently, several experiments and numerical simulations demonstrated remarkable spin-controlled unidirectional coupling between circularly-polarized incident light and transversely-propagating surface or waveguide modes[45–52,125–128], see Figs. 4b,c. In contrast to the spin-directional coupling at metasurfaces (Figs. 3d,e), most of the above experiments dealt with planar interfaces *without* any structures. Moreover, the effect is very robust to the details of the system and equally appears with near 100%-efficiency at metal surfaces[46,47,49], nanofibers[48,50,126], and various waveguides[51,52,125,127,128]. This unique transverse spin-direction coupling originates from fundamental spin properties of *evanescent modes* in free-space Maxwell equations.

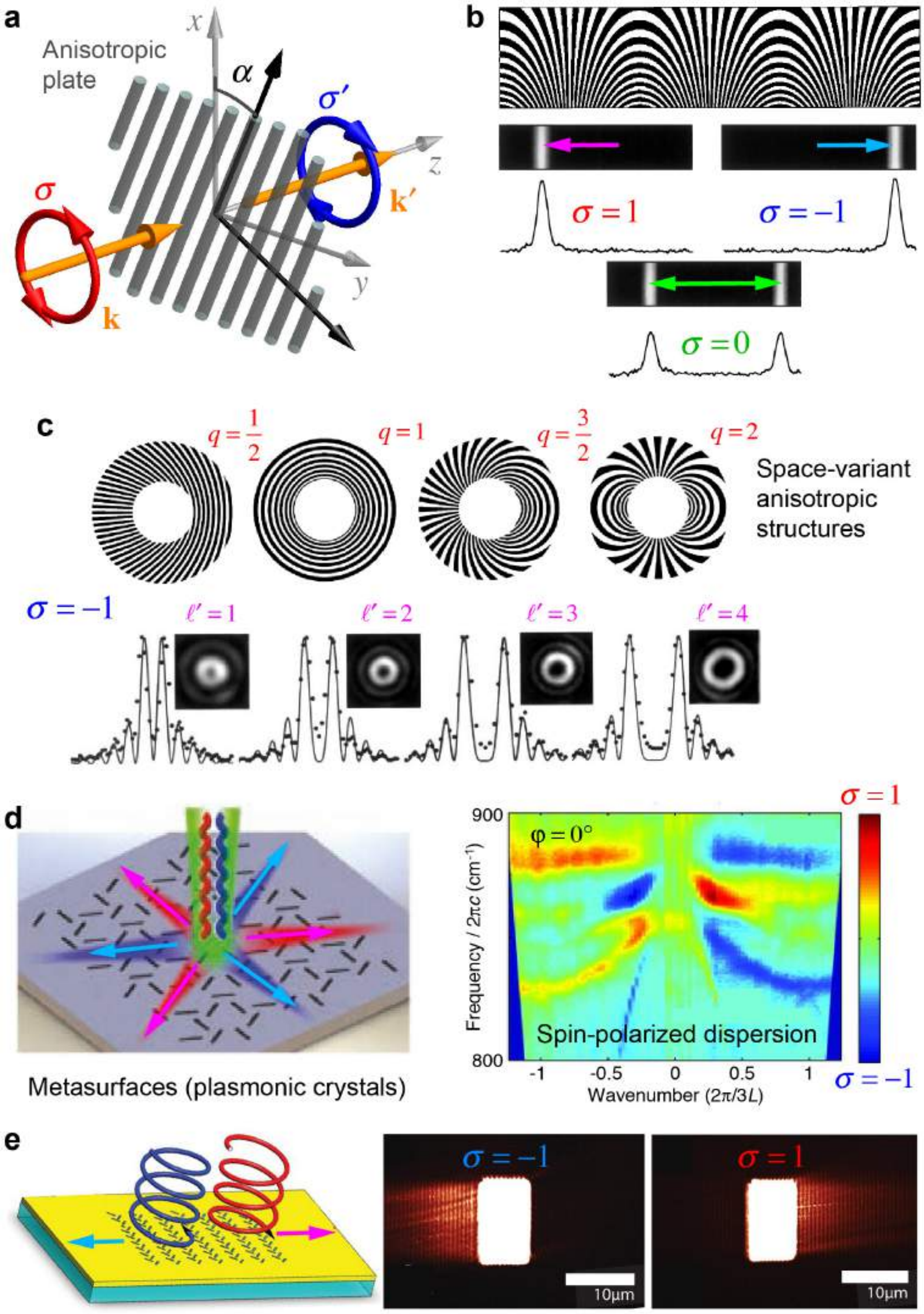

**Figure 3. Spin-orbit interactions induced by planar anisotropic and inhomogeneous structures. (a)** Schematics of light transmission through a locally-anisotropic subwavelength structure with the $\alpha$-oriented anisotropy axis. **(b)** Here the orientation $\alpha$ is linearly-varying with $x$, which induces a helicity-dependent geometric-phase gradient and transverse momentum of the transmitted light. Akin to the spin-Hall effect, this results in the opposite deflection of the two spin states $\sigma = \pm 1$ of light and splitting of the linearly-polarized light ($\sigma = 0$) into two spin components[37]. **(c)** Azimuthally-varying orientation $\alpha = q\varphi$ induces helicity-dependent geometric-phase vortices in the transmitted field[38]. This offers the spin-to-orbital AM conversion and spin-controlled generation of optical vortex beams. **(d)** A periodic plasmonic structure with broken inversion symmetry provides a 2D metamaterial with spin-dependent dispersion analogous to the dispersion of electrons in solids with SOI[44]. **(e)** Such inversion-asymmetric structures provide controllable spin-direction coupling between light and surface plasmons[43].

Until now, we discussed fundamental AM and SOI properties of *propagating* waves (Box 1). Even though some SOI effects were demonstrated in plasmonics system (e.g., Fig. 1e), they just mimicked the propagation properties of usual light. However, evanescent waves can exhibit their unique AM properties. Namely, recently it was discovered that evanescent waves carry extraordinary *transverse spin AM*, which are in sharp contrast to what we knew about the spin of photons[53,54,59].

An evanescent wave, propagating along the $z$-axis and decaying in the $x$-direction, can be regarded as a plane wave with the *complex* wavevector: $\mathbf{k} = k_z \overline{\mathbf{z}} + i\kappa \overline{\mathbf{x}}$, Fig. 4a. Here $k_z > k$ and $k_x = i\kappa$ is the decay constant. Importantly, due to the transversality condition $\mathbf{E} \cdot \mathbf{k} = 0$, which underpins all the intrinsic SOI effects in optics, the evanescent-wave polarization acquires a longitudinal "*imaginary*" (i.e., $\pi/2$ phase-shifted) component: $E_z = -i \dfrac{\kappa}{k_z} E_x$. This means that the electric field of a linearly $x$-polarized wave rotates in the propagation $(x,z)$-plane, and thereby generates a SAM directed along the orthogonal $y$-axis (Fig. 4a). Taking into account both electric- and magnetic-field contributions, it turns out that the transverse spin is independent of the polarization parameters and can be written in a universal vector form:[54,55]

$$\mathbf{S}_\perp = \frac{\mathrm{Re}\,\mathbf{k} \times \mathrm{Im}\,\mathbf{k}}{(\mathrm{Re}\,\mathbf{k})^2}. \tag{6}$$

Transverse SAM (6) represents a completely novel type of optical AM[59], which is in sharp contrast to the usual longitudinal SAM of light, Fig. 1a. Strikingly, it is: (i) orthogonal to the wavevector and (ii) completely independent of the polarization. In particular, the transverse SAM (6) is unrelated to the helicity of light, which is determined by the $(x,y)$ polarization components and associated with the longitudinal $z$-directed SAM. The transverse spin in evanescent waves can be regarded as a distinct manifestation of the SOI of light, which is unrelated to geometric phases and originates from the transversality condition.

Most importantly for applications, the direction of the transverse SAM (6) becomes uniquely locked with the direction of propagation of the evanescent wave. Oppositely-propagating waves with $k_z > 0$ and $k_z < 0$ carry opposite transverse spins $S_y > 0$ and $S_y < 0$, respectively. It is this remarkable feature that is employed in the spin-directional coupling with evanescent waves[45–52,125–128], Figs. 4b,c. Indeed, in all systems[45–52,125–128] the incident light propagated along the transverse $y$-axis carrying usual SAM depending on its helicity: $S_y^{\mathrm{inc}} \propto \sigma$. Then, this incident light was coupled via some scatterer (a nanoparticle, atom, or quantum dot) placed in *evanescent* $x$-decaying tails of the $z$-propagating surface or waveguide modes. Assuming that the SAM of the incident light matches the transverse SAM in the evanescent wave: $S_y^{\mathrm{evan}} \propto \mathrm{sgn}\, k_z$, we find that the propagation direction of the mode with evanescent tails is determined by the helicity of the incident light: $\mathrm{sgn}\, k_z = \sigma$.

Figures 4b,c show two examples of such spin-directional coupling to surface plasmon-polaritons[49] and nanofiber[48] modes. This effect has remarkable 100% efficiency and robustness with respect to the details of the system. It works with *any* interfaces supporting evanescent-tail modes, and offers unique opportunities to be used in spin-chiral networks, spin-controlled gates, and other quantum-optical devices[129].

Remarkably the universal character and robustness of the spin-direction locking in evanescent waves has very deep roots. Recently it has been argued that this phenomenon can be understood as the *quantum spin-Hall effect* of photons[55]. While the spin-Hall effect of light (Fig. 1) originates from the *geometric* phases and Berry curvature for light, the quantum spin-Hall effect is characterized by the *topological* spin-Chern number (spin-weighted integral of the Berry curvature), which is non-zero for the free-space Maxwell equations. This explains the robustness and strong spin-momentum locking, which is quite similar to quantum spin-Hall effect for electrons in topological insulators[130].

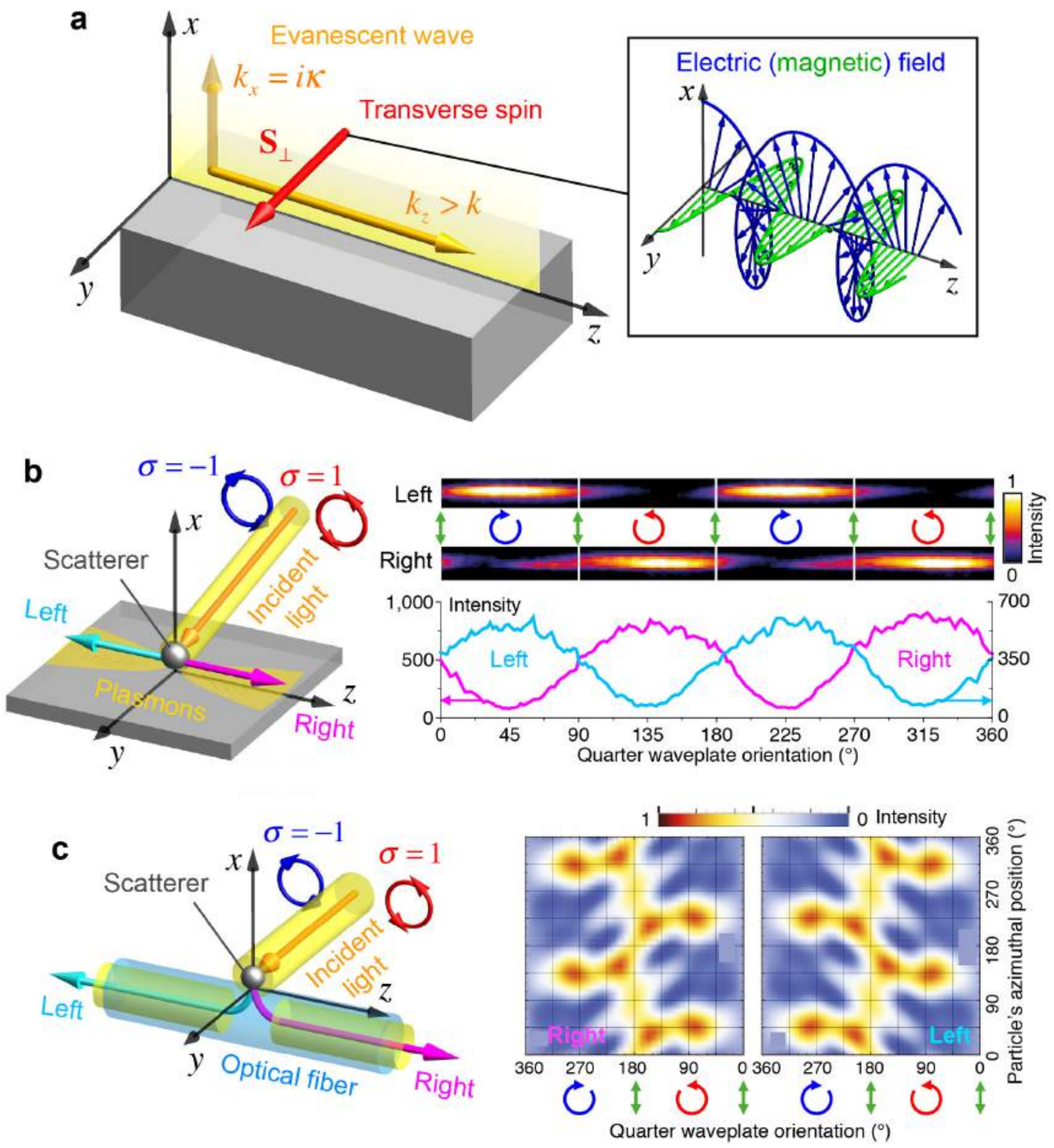

**Figure 4. Transverse spin in evanescent waves and spin-directional interfaces. (a)** A single evanescent wave propagating along the $x = 0$ interface in the $z$-direction and decaying in the $x > 0$ half-space. The complex wavevector $\mathbf{k}$ and the transversality condition generate a $(x, z)$-plane rotation of the wave field (shown in the inset for the linearly $x$-polarized wave) and a transverse $y$-directed spin AM (6) $\mathbf{S}_\perp$. The sign of this spin depends on the propagation direction of the wave (cf., Fig. B1a). **(b)** Spin-controlled unidirectional coupling of the $y$-propagating light to the $z$-propagating surface plasmon-polaritons[49]. The spin AM of the incident field matches the transverse spin of the surface plasmon and determines its direction of propagation. **(c)** An analogous transverse spin-direction coupling occurs for the $z$-propagating modes of an optical fiber[48]. These modes are coupled to the $y$-propagating light via the same evanescent tails as surface plasmons in (b).

## Concluding remarks

We have shown that the spin-orbit interactions of light originate from fundamental properties of the electromagnetic Maxwell waves and are inherent in all basic optical processes. Like relativistic SOI of electrons, optical SOI effects are typically small in geometrical-optics processes dealing with scales and structures much larger than the wavelength. However, these phenomena become important and crucially determine the behavior of light at subwavelength scales and structures of modern nano-optics, photonics, and plasmonics. This is why the SOI of light are currently attracting rapidly growing interest.

The SOI of light have both fundamental and applied importance for physics. On the one hand, these phenomena allow the direct observation of fundamental spin-induced effects in the dynamics of relativistic spinning particles (photons). Measurements of similar effects, e.g., for Dirac electrons or for analogous condensed-matter quasi-particles, are far beyond current capabilities. On the other hand, akin to significant enhancement of electron SOI in solid-state crystals, the SOI of light are considerably enhanced and can be artificially designed in optical nanostructures, including metamaterials. This paves the way to *spinoptics*: an optical counterpart of electron spintronics in solids. Introducing additional spin degrees of freedom for smart control of light promises highly important applications in photonics, optical communications, metrology, and quantum information processing. In this manner, the SOI of light conform to the most important trends in modern engineering: (i) miniaturization of devices down to subwavelength scales and (ii) increasing amount of information due to additional internal degrees of freedom.

Examples shown throughout this review clearly show that SOI phenomena can play diverse roles in various optical setups. On the one hand, they are inevitably present as small wavelength-scale aberrations in *any* optical interfaces and lenses (Figures 1 and 2). Thus, these effects must be taken into account in any precision devices and measurements. On the other hand, they can dramatically affect and control the intensity and propagation of light in tightly-focused fields and structured media. Moreover, since SOI phenomena are usually determined by basic symmetry properties and are robust with respect to perturbations in the system, it is natural to employ these phenomena for spin-dependent shaping and control of light. In particular, Figures 2–4 show examples of the following fully spin-controlled, near-100% efficient, and robust processes: (i) optical manipulation of small particles, (ii) zero-to-maximum intensity switching, (iii) subwavelength optical probing of nanoparticles, (iv) directional propagation and diffraction, (v) generation of vortex beams, (vi) propagation and spectrum of Bloch modes in metamaterials, and (vii) unidirectional excitation of surface and waveguide modes.

Thus, the spin-orbit interactions of light represent an important and integral part of modern optics. We hope that this review will aid further progress in this rapidly advancing area by forming an effective framework for future studies and applications of optical spin-orbit phenomena.

## Acknowledgements

This work was supported, in part, by the RIKEN iTHES Project, MURI Center for Dynamic Magneto-Optics, JSPS-RFBR contract no. 12-02-92100, Grant-in-Aid for Scientific Research (S), EPSRC (UK) and the ERC iPLASMM project (321268). A.Z. acknowledges support from the Royal Society and the Wolfson Foundation.